# Cloaking of Matter Waves

**Shuang Zhang, Dentcho A. Genov, Cheng Sun and Xiang Zhang**[*]

*5130 Etcheverry Hall, Nanoscale Science and Engineering Center, University of California, Berkeley, California 94720-1740, USA*

Invariant transformation for quantum mechanical systems is proposed. A cloaking of matter wave can be realized at given energy by designing the potential and effective mass of the matter waves in the cloaking region. The general conditions required for such a cloaking are determined and confirmed by both the wave and particle (classical) approaches. We show that it may be possible to construct such a cloaking system for cold atoms using optical lattices.

The advancement of plasmonic and metamaterial physics has enabled the realization of new realm of optics, such as extraordinary optical transmission through an array of subwavelength apertures [1, 2], a superlens [3, 4] that overcomes the diffraction limits, and a cloaking devices that can hide an object from external electromagnetic radiation. [5, 6, 7] Following the recent theoretical works by Pendry [5] and Leonhard [6], electromagnetic cloaking has been intensively studied, with the first experimental demonstration at microwave frequency accomplished by Schurig et al [7]. Wave cloaking in elastometric system has also been studied theoretically, with Milton's claim of invariant transformation of elastodynamic wave under the limitation of harmonic mappings [8]. On the other hand, Cummer et al demonstrated the equivalence between electrodynamics and elastodynamics in the two dimensional case [9]. Recently, an electron focusing effect across a p-n junction in Graphene film, that mimics the Veselago's lens in optics, has been proposed. [10] This, as well as the theoretical demonstration of 100% transmission of cold rubidium atom through an array of sub de Broglie wavelength slits, brings the original continuous wave phenomenon in contact with the quantum world. [11]

---

[*] Corresponding author.   Email: xiang@berkeley.edu

Cloaking of electromagnetic waves is possible due to time invariant coordinate transformation of the governing Maxwell's equations. Such invariant transformations map a particular region in free space to a spatial domain with position dependent and anisotropic material parameters (such as permeability and permittivity in electromagnetics). In this letter, we study an invariant transformation of the Schrödinger equation for quantum waves and show that matter cloaking can be achieved by spatially controlling the potential and effective mass of a particle as it travels inside the media. The extension of the invariant coordinate transformation to quantum mechanical systems could open a new field of study since it promises unprecedented control of the quantum wave, and may lead to new phenomena and applications.

Effective mass Schrödinger equations are widely used for systems with spatially varying material properties [12,13,14]. By taking into account material anisotropy, the time independent Schrödinger equation is written as:

$$-\frac{\hbar^2}{2m_0}\vec{\nabla}\cdot(\hat{m}^{*-1}\vec{\nabla}\psi) + V\psi = E\psi \qquad (1)$$

where the spatially dependent and anisotropic effective mass $\hat{m}^* = m_0\hat{m}$ is generally a tensor ($m_0$ is the mass in free space), and $V(\vec{r})$ is a "macroscopic" potential. For instance, for electrons in a crystal with slowly varying composition, $V(\vec{r}) = E_b(\vec{r}) + U(\vec{r})$, where $E_b(\vec{r})$ is the energy of the local band edge and $U(\vec{r})$ is a slowly varying external potential [15]. The above equation can also be rewritten as two first order differential equations

$$\vec{u} = \hat{m}^{-1}\vec{\nabla}\psi, \quad -\frac{\hbar^2}{2m_0}\vec{\nabla}\cdot\vec{u} = (E-V)\psi \qquad (2)$$

Utilizing the form Eq. 2, we consider an invariant co-ordinate transformation $(x_1, x_2, x_3) \rightarrow (q_1, q_2, q_3)$, by assuming both co-ordinate bases to be orthogonal. It is straightforward to show that divergence of vector $\vec{u}$ and gradient of the wave function $\psi$ in the old coordinate frame are related to those in the new coordinates by

$$\vec{\nabla}_{\vec{x}}\psi = \hat{h}^{-1}\vec{\nabla}_{\vec{q}}\psi, \quad \vec{\nabla}_{\vec{x}}\cdot\vec{u} = \frac{1}{|\det(\hat{h})|}\vec{\nabla}_{\vec{q}}\cdot\vec{v} \qquad (3)$$

where $h_i = |\partial\vec{x}/\partial q_i|$ are the Lamé coefficients, $\hat{h}_{ij} = h_i\delta_{ij}$ ($\delta_{ij}$ is the Kronecker delta), and we define a new vector $\vec{v} = |\det(\hat{h})|\hat{h}^{-1}\vec{u}$. Combining Eq. 2 and 3 we thus obtain the Schrödinger equations in the new co-ordinate system,

$$-\frac{\hbar^2}{2m_0}\nabla_q\cdot\vec{v} = \det|\hat{h}|(E-V)\psi \qquad (4)$$
$$\vec{v} = \det|\hat{h}|(\hat{h}\hat{m}\hat{h})^{-1}\vec{\nabla}_q\psi$$

Clearly, Eqs. 4 are mathematically equivalent to the Eq. 2, under the following transformations:

$$\hat{m}' = \frac{\hat{h}\hat{m}\hat{h}}{\det|\hat{h}|}, \quad V' = E + |\det(\hat{h})|(V-E) \qquad (5)$$

of the potential and effective mass, respectively. Those relationships constitute the general conditions for matter wave cloaking. Since the Equations (5) depends on the energy (frequency) of the quantum wave, it indicates that invariant transformation does not exist for the general time dependent case.

As an example of the proposed quantum mechanical transformation we study the cloaking of a quantum wave in a spherically symmetric system, where the object to be hided is to be contained inside a sphere of radius $r_1$, and the cloaking medium comprises a spherical shell from $r_1$ to an outer radius $r_2$. The relation between the old and new coordinate is given by $r' = g(r)$, $\theta' = \theta$ and $\phi' = \phi$, where $g(r)$ is a monotonic radial scaling function with $g(r_1) = 0$ and $g(r_2) = r_2$. Without any loss of generality we further assume that the potential outside the cloaking region is zero. According to equations (5) the potential and mass in the cloaking shell are given as

$$\hat{V}(\vec{r}, E) = [1 - (\frac{g}{r})^2 g'(r)] E$$
$$m_{rr} = g'(r)(\frac{r}{g})^2 m_0, \quad m_{\theta\theta} = m_{\phi\phi} = \frac{1}{g'(r)} m_0$$
(6)

where we also assume a free space propagation in the original co-ordinate space ($V = 0$, $\hat{m}_{ij}^* = m_0 \delta_{ij}$). Now if we consider an arbitrary wave incident on the cloaking system, the wave inside the cloaking region could be expanded into spherical harmonics as $\psi_c(r,\theta,\phi) = \sum_{lm} c_{lm} f_l(r) Y_l^m(\theta,\phi)$. By combining the Schrodinger Eqn (1) and cloaking conditions Eqn. (6), we obtain that the radial function $f_l(r)$ has a solution in the form of spherical Bessel functions of the first kind $f(r) = j_l(k_0 g(r))$, where $k_0 = \sqrt{2mE}/\hbar$. The incident and scattered waves outside the cloaking region are also expanded into spherical harmonics as $\psi_i = \sum_{lm} a_{lm} j_l(k_0 r) Y_{lm}(\theta,\phi)$ and $\psi_s = \sum_{lm} b_{lm} h_l(k_0 r) Y_{lm}(\theta,\phi)$ respectively, where $h_l$ are the spherical Hankel functions. From the continuity condition $\psi_i + \psi_s = \psi_c$ and conservation of current $m_0^{-1}(\partial_r \psi_i + \partial_r \psi_s) = m_{rr}^{-1} \partial_r \psi_c$ at the outer surface $r = r_2$ we also obtain

$$a_{lm} j_l(k_0 r_2) + b_{lm} h_l(k_0 r_2) = c_{lm} j_l(k_0 r_2)$$
$$a_{lm} j_l'(k_0 r_2) + b_{lm} h_l'(k_0 r_2) = \frac{m_0}{m_{rr}} g'(r) c_{lm} j_l'(k_0 r_2) = c_{lm} k_0 j_l'(k_0 r_2)$$
(7)

It immediately follows from Eqn (7) that $b_{lm} = 0$ and $c_{lm} = a_{lm}$, namely, the incident quantum wave propagates throughout the cloak without any scattering and distortion. In addition, the wave inside the cloak is $\psi_c(r,\theta,\phi) = \sum_{lm} a_{lm} k_0 j_l(g(r)) Y_l^m(\theta,\phi) = \psi_i^0(g(r),\theta,\phi)$, where $\psi_i^0$ is what the wave function of the incident wave would be for $r<r_2$ if both the cloak and object were absent. This directly verifies the conformal transformations Eqs. (5 and 6) and confirms a cloaking of an incident wave at the designed energy level.

When the energy of the incident wave deviates from the designed cloaking energy, the matter wave will suffer distortion and scattering as it passes through the system. To look into this issue, we resort to the classical limit and investigate the particle trajectories under the configuration shown in Fig. 1. In the case of a linear scaling function $g(r) = (r - r_1)/\eta$, where $\eta = (r_2 - r_1)/r_2$, the trajectories can be analytically calculated for arbitrary particle energy, thus giving us some useful insight on how critical the cloaking condition needs to be satisfied. Inside the cloaking region, the Hamiltonian is written as,

$$H = \frac{P_r^2}{2m_{rr}} + \frac{P_\theta^2}{2m_{\theta\theta} r^2} + \frac{P_\phi^2}{2m_{\phi\phi} r^2 \sin^2\theta} + \hat{V}(r, E_0) \qquad (8)$$

and the impinging particle is assumed to have an energy $E$ that does not necessarily coincide with the designed cloaking energy $E_0$. Due to the spherical symmetry there is a freedom to choose the plane of motion which we set to be the $\theta = \pi/2$ plane (equatorial motion). In this case, $P_\theta = 0$ and $P_\phi$ are constants of motion and the trajectory inside the cloak is calculated from the Hamiltonian as,

$$\frac{d\phi}{dr} = \frac{\partial H/\partial P_\phi}{\partial H/\partial P_r} = \pm \frac{1}{(r-r_1)\sqrt{A(r-r_1)^2 + B(r-r_1) - C}} \qquad (9)$$

where $A = (E_0 + \eta^3 \Delta E)/\eta^2 b^2 E$, $B = 2r_1 \eta \Delta E / b^2 E$, $C = 1 - r_1 B/2$, and $b$ is the impact

parameter. (see Fig. 1). The sign on the right hand side of Eqn (9) is negative for the particle approaching the inner surface of the cloaking sphere and vice versa. The integration of equation (9) gives a range of particle trajectories, as presented in Fig. 2. For the unique case $E = E_0$ ($\Delta E = 0$), presented in Fig. 2 (a), all particles follows conformal paths, leaving the cloaking region with precisely the same momentums and positions as if the scattering region was absent. A matter wave (or particles) with incident energy different from the designed energy will suffer a level of distortion as it passes through the system, with a deflection angle given by $\alpha = 2\phi_0 + \frac{2}{\sqrt{C}} \cos^{-1}(\frac{2C/(r_2 - r_1) - B}{\sqrt{B^2 + 4AC}})$. This effect is shown in Figs. 2b, and 2c where we show that incident particle energies with a slight energy deviation of 10% leads to substantial change in the particle trajectories, especially at small impact parameters.

One natural question to ask is what type of system can be used for the implementation of the proposed matter wave cloaking. Here we consider the cold atoms in an optical lattice. It has been shown that effective mass of cold atoms can be significantly increased along the direction of an optical potential modulation [16, 17, 18, 19, 20]. Atoms inside a one dimensional optical lattice experience effective potential of the form $V_{op} = sE_R \sin^2(k_{op}x)$, where $s$ is a dimensionless parameter indicating the amplitude of the optical potential, $E_R$ is the recoil energy given by $\hbar^2 k_{op}^2 / 2m_0$ and $k_{op}$ is the wave vector of light. The effective mass of the atoms $m^*$ and band edge energy $E_b$ can be obtained by solving the Mathieu's equation [19], and are plotted in Fig. 3(a).

In the proposed cloaking system, the cloaking shell consists of optical standing waves with slowly varying amplitude along the radial direction in combination with a carefully designed external magnetic potential. By designing the optical intensity locally, the effective mass can be engineered to satisfy the cloaking requirement for $m_r$. While along the angular directions, there is no potential modulation and thus the mass remains $m_0$, which does not satisfy the cloaking requirement Eqn (6). To solve this problem, we resort to cloaking in a reduced form with perfect impedance matching at the outer shell of the cloak [21]. In the classical limit, for a particle traveling through two cloaking systems with different sets of parameters ($\hat{m}_1, V_1$) and ($\hat{m}_2, V_2$) with trajectories $r_1(t), \theta_1(t), \phi_1(t)$ and $r_2(t), \theta_2(t), \phi_2(t)$, if the initial positions and momenta are the same and $\dot{r}_1/\dot{\theta}_1 = \dot{r}_2/\dot{\theta}_2$ and $\dot{r}_1/\dot{\phi}_1 = \dot{r}_2/\dot{\phi}_2$ are satisfied everywhere inside the cloaking region, then the two sets of trajectories would exactly match. Using the Hamiltonian Eqn (8), it is straightforward to show that this leads to the relations $m_{r1}/m_{r2} = m_{\theta 1}/m_{\theta 2} = m_{\phi 1}/m_{\phi 2}$ and $V_2 = (1 - m_{r1}/m_{r2})E + (m_{r1}/m_{r2})V_1$. If ($\hat{m}_1, V_1$) correspond to the perfect cloaking conditions Eqn (6), then ($\hat{m}_2, V_2$) would be a set of reduced cloaking conditions. Thus, in a reduced cloaking system consisting of concentric optical lattices, we have

$$m_{\theta 2} = m_{\phi 2} = m_0, \quad m_{r2} = m_0 (rg'(r)/g)^2$$
$$V_2 = [1 - (g(r)/r)^2]E \quad (10)$$

Furthermore, as shown in [21], the reflection at the outer shell is eliminated if a nonlinear radial scaling function $g(r)$ is chosen such that $g'(r)|_{r=r_2} = 1$, i.e. $m_{r2} = m_0$ and $V_2 = 0$ at the outer shell of the cloak.

Eqns (10) shows that $m_{r2}$ approaches infinity at the inner shell of the cloak. This would not be possible in a realistic cloaking design using optical lattices. However, given the largest effective mass $\overline{m}_r$ that can be achieved, a truncated radial effective mass profile $m_r(r)$ may suffice to effectively reduce the scattering cross section and achieve a reasonable level of cloaking. To show that we consider a design where $m_r$ linearly increases from $m_0$ at $r = r_2$ to $\overline{m}_r$ at $r = r_2 - \delta$, slowly enough to ensure negligible reflection. The effective mass $m_r$ is then kept constant at $\overline{m}_r$ from $r = r_2 - \delta$ to $r = r_1$. With this profile for $m_r$ and using Eqn (10), we solve for $g(r)$ and $V(r)$ [Fig. 3(b)]. Although $g(r_1) \neq 0$ as in the perfect cloaking case, still, it can be significantly less than the object radius $r_1$, indicating a much decreased scattering cross section $\sigma_s = \pi g^2(r_1)$ compared to that without cloaking.

As a particular material system we rely on Ref [20], and design the system to cloak ultra-cold Rb in an optical lattice with period 427nm. The atoms quantum wavelength is set at 5μm or 10 times longer than the lattice period which allows for the optical lattice to be treated as an effective medium. The chosen de Broglie wavelength corresponds to a temperature close to 4.5 nK, which is about one order of magnitude higher than current experiment limit of 0.5 nK [22]. The maximum optical lattice depth is set to 12.8 $E_R$, which is less than 18.5$E_R$ that has been already reported in the literature [20]. According to Fig. 3(a), this optical lattice depth could provide an effective mass of $\overline{m}_r = 10 m_0$ along the radial direction. In addition, carefully designed magnetic field is used to achieve the desired slowly-varying effective potential profile. It has been shown that almost arbitrary potential patterns could be generated using micro magnetic traps [23]. Based on the proposed material system and for $r_1 = 100\,\mu m$, $r_2 = 400\,\mu m$ and $\delta = 100\,\mu m$, Eqns (10) gives $g(r_1) = 6.4\,\mu m$, which corresponds to a reduction of scattering cross section by a factor of 250. The combined optical and magnetic potential $V_c$ that generate the desired cloaking parameter profiles are shown in Fig. 3 (c). We note that the amplitude of $V_c$ is more than two orders of magnitude larger than the kinetic energy of the incident atom, which would pose some technical challenges in realizing the system. However, there is no fundamental limit for obtaining such accuracy in the optically and magnetically induced potential. Finally, one may ask how a concentric optical lattice can be achieved. A 3D spherical concentric optical lattice might be difficult to construct, however, some recent work has been done to generate complicated 2D optical patterns including concentric optical rings for the application of steering or trapping atoms [24, 25, 26], which may be used to achieve a 2D cloaking system for cold atoms.

In conclusion, we have shown that the time independent Schrodinger equations can be invariantly transformed to achieve cloaking of matter waves. We verified this result by calculating both the scattering of quantum waves incident upon the cloak and the trajectories of classical particles inside such cloaking system. Finally, we proposed a possible scheme to achieve cloaking of cold atoms using concentric optical lattices. This work might be potentially important for the controlling of electrons in inhomogeneous crystal systems, cold atoms in an optical lattices, radiation shielding, particle beam steering.

This work was supported by AFOSR MURI (Grant No. 50432), SINAM and NSEC under Grant No. DMI-0327077.

Figure captions:

**Fig. 1** (color online) Schematic of a spherical cloaking system with interaction potential and effective mass tuned in the shell region between $r_1$ and $r_2$. An impinging particle travels along +y direction and has an initial momentum $P_0$, and impact parameter $b$. After passing throughout the system it leaves at an angle $\alpha$.

**Fig. 2** The trajectories of particles passing through a cloaking system with $r_2 = 2r_1$ for (a) $\Delta E = 0$ (b) $\Delta E = -0.1E_0$, and (b) $\Delta E = 0.1E_0$ .

**Fig. 3** (a) The dependence of effective mass and band edge energy on the amplitude of the optical lattice potential $s$ (b) The plot of the radial scaling function $g$ (black), effective mass along radial direction $m_r$ (gray dashed) and potential $V$ (gray) for a reduced cloaking design with $r_2 = 4r_1$, $\bar{m} = 10m_0$ and $\delta = r_1$. (c) The combined optical and magnetic potential profile to achieve the proposed cloaking system for $r_1$=100 μm, $r_2$=400 μm and δ=100 μm. Inset is a magnified view of the shadowed area.

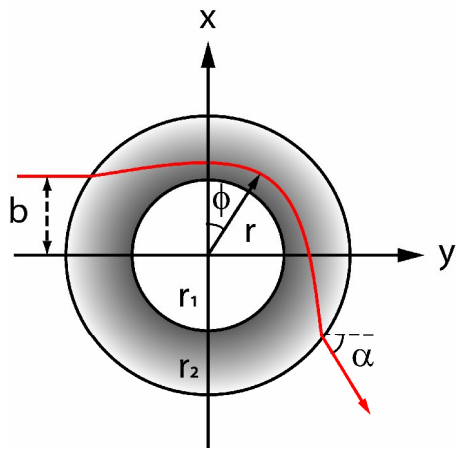

Fig. 1

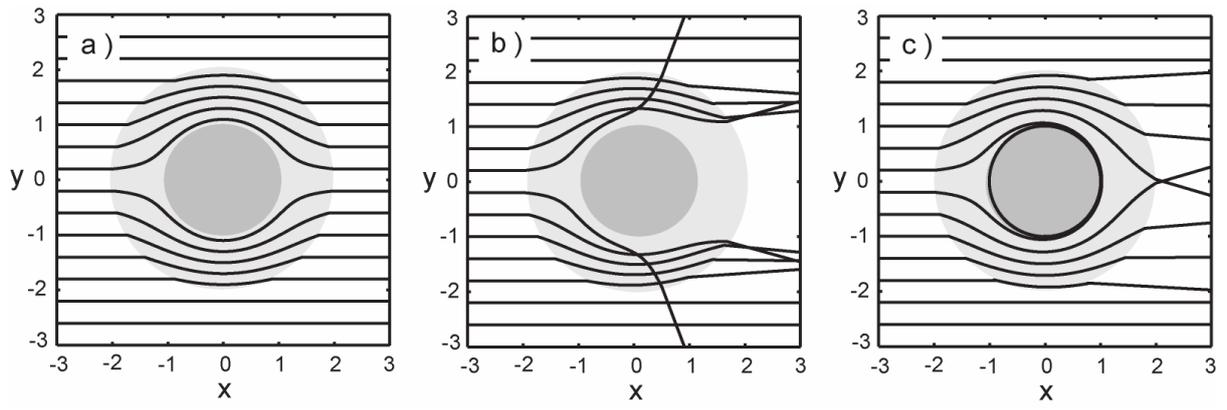

Fig. 2

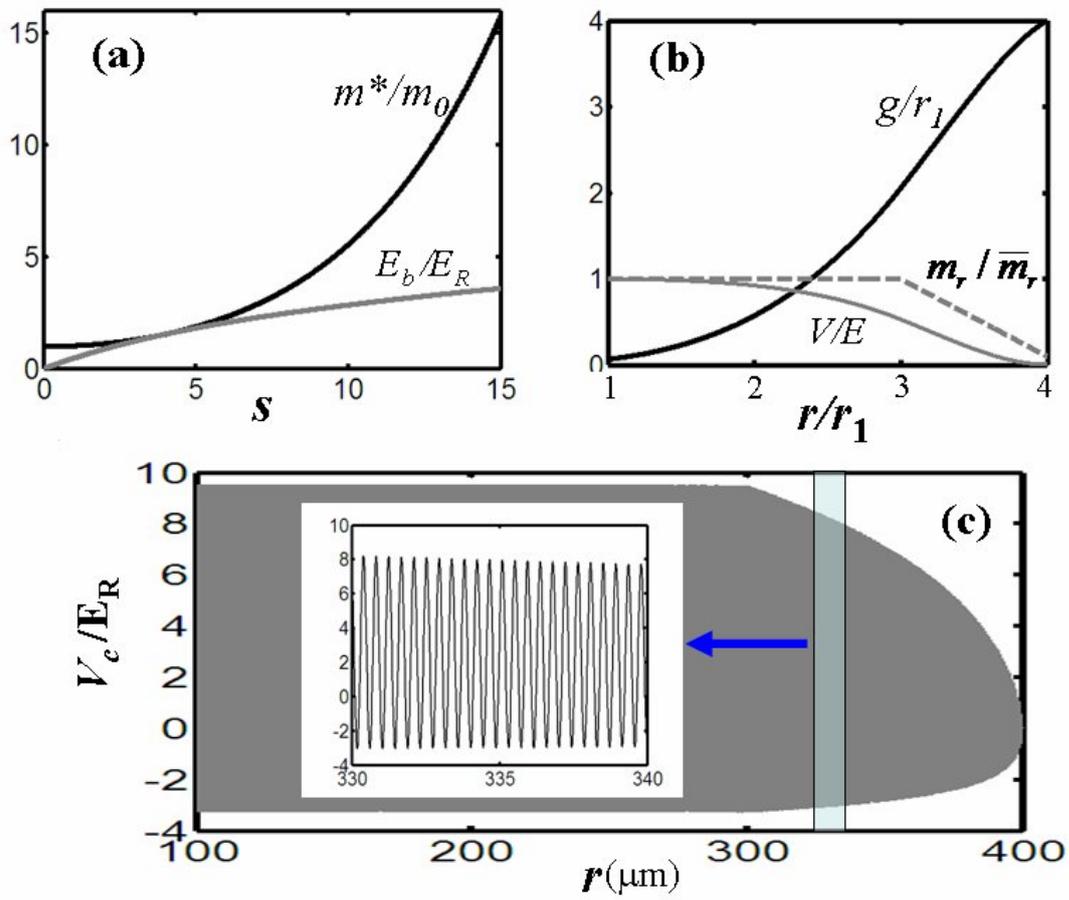

Fig. 3